%% file: main.tex
\documentclass{vgtc}                          

\graphicspath{{figures/}{pictures/}{images/}{./}} 

\usepackage{times}                     

\usepackage{tabu}                      
\usepackage{booktabs}                  
\usepackage{lipsum}                    
\usepackage{mwe}                       

\usepackage{mathptmx}                  

\onlineid{0}

\vgtccategory{Research}

\vgtcinsertpkg

\title{Exploring Organizational Strategies \\ in Immersive Computational Notebooks}

\author{Sungwon In\thanks{e-mail: sungwoni@vt.edu}\\ %
        \scriptsize Virginia Tech %
\and Ayush Roy\thanks{e-mail: ayushroy24@vt.edu}\\ %
     \scriptsize Virginia Tech %
\and Eric Krokos\thanks{e-mail: ericpkrokos@gmail.com}\\ %
     \scriptsize Department of Defense %
\and Kirsten Whitley\thanks{e-mail: visual.tycho@gmail.com}\\ %
     \scriptsize Department of Defense %
\and Chris North\thanks{e-mail: chnorth1@vt.edu}\\ %
     \scriptsize Virginia Tech %
\and Yalong Yang\thanks{e-mail: yalong.yang@gatech.edu}\\ %
     \scriptsize Georgia Tech}

\abstract{
Computational notebooks, which integrate code, documentation, tags, and visualizations into a single document, have become increasingly popular for data analysis tasks. 
With the advent of immersive technologies, these notebooks have evolved into a new paradigm, enabling more interactive and intuitive ways to perform data analysis.
An immersive computational notebook, which integrates computational notebooks within an immersive environment, significantly enhances navigation performance with embodied interactions. 
However, despite recognizing the significance of organizational strategies in the immersive data science process, the organizational strategies for using immersive notebooks remain largely unexplored.
In response, our research aims to deepen our understanding of organizations, especially focusing on spatial structures for computational notebooks, and to examine how various execution orders can be visualized in an immersive context.
Through an exploratory user study, we found participants preferred organizing notebooks in half-cylindrical structures and engaged significantly more in non-linear analysis.
Notably, as the scale of the notebooks increased (i.e., more code cells), users increasingly adopted multiple, concurrent non-linear analytical approaches.
}

\keywords{Computational Notebook, Immersive Analytics, Virtual Reality, Data Science, 3D UI, Space Organization Strategies}

\teaser{
    \centering
    \includegraphics[width=0.95\linewidth]{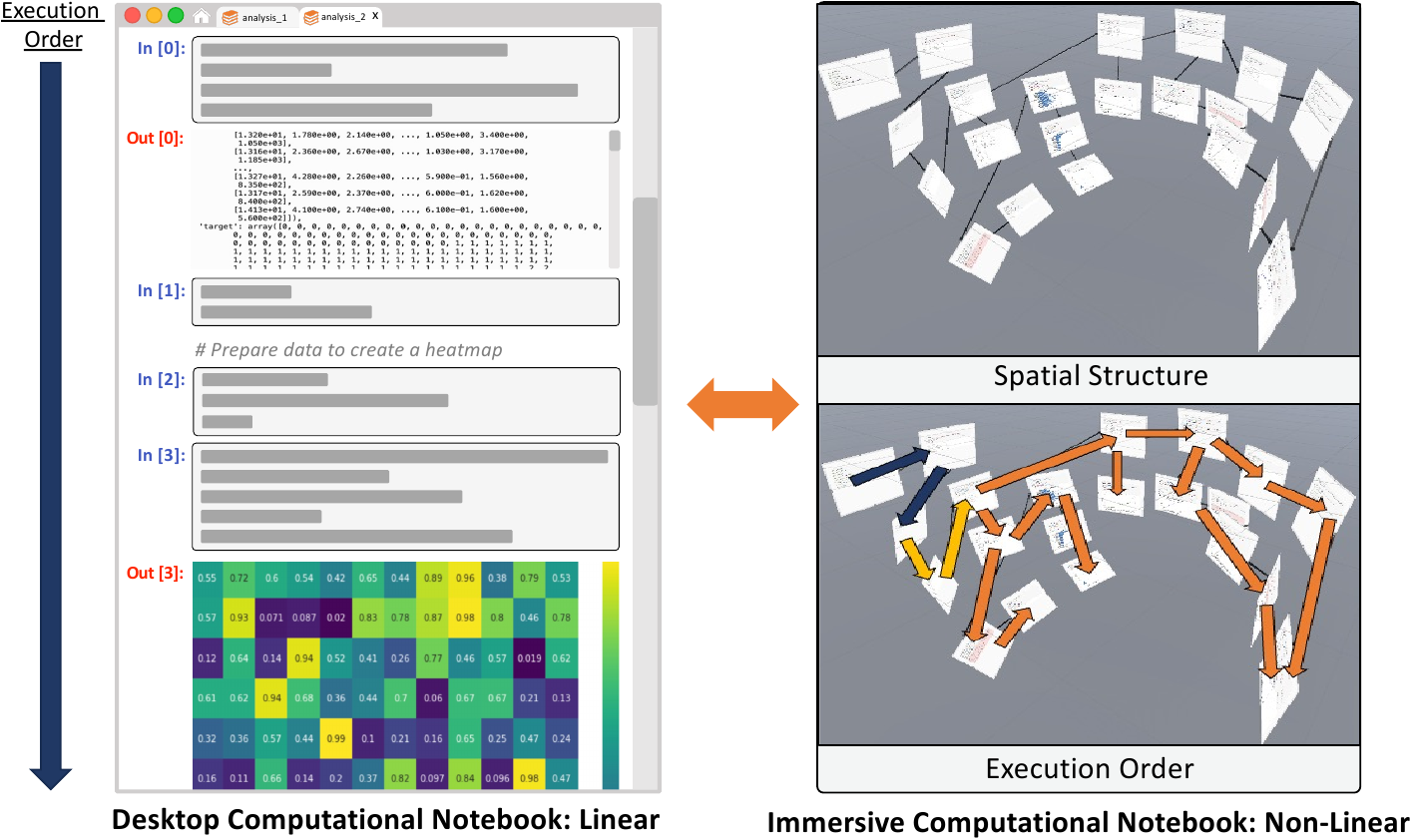}
    \caption{Cells are displayed linearly on a desktop computational notebook, resulting in a linear execution order (left). However, an immersive computational notebook enables various types of non-linear execution orders (right).}
    \label{fig:teaser}
}

\usepackage{tabu}                      
\usepackage{booktabs}                  
\usepackage{lipsum}                    
\usepackage{mwe}                       

\usepackage{hyperref}
\usepackage[english]{babel}
\addto\extrasenglish{%
}

\AtBeginDocument{%
  }

\usepackage{subfigure}
\usepackage{makecell}

\usepackage{enumitem}
\setitemize{noitemsep,topsep=0pt,parsep=0pt,partopsep=0pt}

\usepackage{hyphenat}

\usepackage{forloop}

\usepackage[defaultcolor=black]{changes}

\newsavebox\MyBreakChar%
\sbox\MyBreakChar{}
\newsavebox\MySpaceBreakChar%
\sbox\MySpaceBreakChar{\hyp}
\makeatletter%
\newcommand*{\BreakableChar}[1][\MyBreakChar]{%
  \leavevmode%
  \discretionary{\usebox#1}{}{}%
}%
\makeatother

\newcounter{index}%
\newcommand{\AddBreakableChars}[1]{%
  \StrLen{#1 }[\stringLength]%
  \forloop[1]{index}{1}{\value{index}<\stringLength}{%
    \StrChar{#1}{\value{index}}[\currentLetter]%
    \IfStrEqCase{\currentLetter}{%
        {*}{\currentLetter\BreakableChar[\MyBreakChar]}%
        {/}{\currentLetter\BreakableChar[\MyBreakChar]}%
        {+}{\currentLetter\BreakableChar[\MyBreakChar]}%
        {\&}{\currentLetter\BreakableChar[\MyBreakChar]}%
    }[\currentLetter]%
  }%
}%

\def\simple{$\vcenter{\hbox{\includegraphics[height=0.93em]{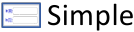}}}$}
\def\complex{$\vcenter{\hbox{\includegraphics[height=0.97em]{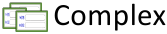}}}$}

\def\simpleB{$\vcenter{\hbox{\includegraphics[height=1em]{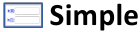}}}$}
\def\complexB{$\vcenter{\hbox{\includegraphics[height=1em]{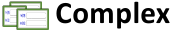}}}$}

\begin{document}

\maketitle

\input{body/01-Introduction-v2}

\input{body/02-Related_Work}

\input{body/03-User_Study}

\input{body/04-Results}

\input{body/05-Discussion}
\input{body/06-Implication}
\input{body/07-Limitation}
\input{body/08-Conclusion}


\acknowledgments{
This research was supported by industry, government, and institute members of the NSF SHREC Center, which was founded in the IUCRC program of the National Science Foundation, and NSF grant IIS-2441310.
}

\bibliographystyle{abbrv-doi}
\bibliography{main}

\end{document}

%% file: body/01-Introduction-v2.tex
\section{Introduction}
\label{sec:intro}

Computational notebooks are structured as a series of ``cells'' that integrate code, documentation, tags, and visualizations into a single document~\cite{Kluyver2016jupyter}, streamlining complex data analysis tasks such as building analytical pipelines, debugging, and comparative analysis~\cite{chattopadhyay2020s, Kluyver2016jupyter}. 
This integration has introduced a new paradigm in data science, making them a popular choice among data analysts~\cite{JupyterSurvey}.
However, as data analysis grows in complexity, desktop-based computational notebooks struggle to manage the messiness of cells due to their fixed structures, making it difficult to reorganize workflows and navigate efficiently~\cite{head2019managing}.

Execution order, a unique characteristic of computational notebooks, is inherently related to the organization and navigation of cells, as it determines how analytical processes are structured and executed. 
In desktop-based notebooks, the execution order is typically linear, represented from top to bottom~\cite{kery2018story} (see \autoref{fig:teaser}). 
However, when given greater flexibility in organization, analysts often adopt non-linear structures~\cite{elmqvist2008melange, harden2022exploring, kang2021toonnote}, as linearity can complicate sense-making in real-world analytical workflows~\cite{deline_glinda_2021}. 
To further support non-linearity, In et al. introduced the Immersive Computational Notebook~\cite{in2024evaluating}. 
Their findings suggest that embodied interactions are more intuitive and effective in building non-linear workflows.
However, as the analysis and sense-making process became more complex, relying solely on an initial half-cylindrical structure made it more challenging to develop the structure in a way that made sense to them, such as incorporating hierarchical structures and depth cues.
While half-cylindrical structures are effective for navigating digital contents within immersive environments~\cite{andrews2010space, liu2020design, liu2022effects}, strategies for structuring spatial elements and execution order to support sense-making within immersive computational notebooks have not been explored yet.
\textbf{Therefore, our first goal is to identify various types of organizational strategies within immersive computational notebooks, specifically focusing on how spatial elements and execution order can be organized in immersive spaces.}

Building on our primary goal of identifying organizational strategies for immersive computational notebooks, the number of code cells can significantly influence the process of shaping these cells.
As the number of code cells increases within immersive environments, the organization expands its size to accommodate the growing number of code cells.
For instance, users may expand their organization vertically by placing cells more on the top or bottom to avoid horizontal expansion, resulting in excessive head movements that can lead to physical discomfort and strain~\cite{tseng2023understanding}.
Consequently, optimizing interactions may drive users to adopt different organizational strategies to manage their workflow more effectively.
\textbf{To this end, our second goal is to explore how the increasing number of code cells impacts their organizations within immersive environments.}

To systematically investigate our proposed goals, we conducted an exploratory user study under two conditions: the ``Simple'', which involved fewer code cells to mimic basic data analysis tasks, and the ``Complex'', with more code cells representing complex data analysis tasks. 
In the initial setup, we provided participants with a randomly ordered, piled-up set of cells, intentionally avoiding predefined structures to prevent bias in their organizational strategies.
Each cell contained a single executed code snippet with either visual outputs or data previews.
Participants were tasked with organizing the cells in a manner that made sense to them. 
Our results indicated a preference for half-cylindrical structures, although quarter-cylindrical structures were widely adopted.
\added{Notably, participants creatively and actively engaged in non-linear analysis workflows.}
In addition, as the number of code cells increased, participants experienced their structures' growth in size, with an increasing number adopting multiple execution orders.

%% file: body/02-Related_Work.tex

\section{Related Work}
\label{sec:relwork}





\textbf{Computational Notebooks.}
Computational notebooks are a widely adopted data science tool that facilitates incremental and iterative analysis by integrating text, visualizations, markdown, and code cells within a single document~\cite{Kluyver2016jupyter}. 
However, Chattopadhyay et al. identified analysts' struggles with computational notebooks, particularly during the iterative process of exploratory data analysis~\cite{chattopadhyay2020s}. 
A notable challenge is the inability to perform nonlinear analysis, such as comparing code blocks or visualizations that are not sequentially aligned. 
This limitation is largely due to the linear nature of traditional computational notebooks, which hampers analysis and navigation process~\cite{deline2012debugger}.
In response, In et al. explored using computational notebooks within immersive environments~\cite{in2024evaluating}. 
They specifically evaluated the potential of physical navigation and a branch-and-merge mechanism to facilitate non-linearity.
Their findings indicated that physical movement offers effective navigation, and leveraging non-linearity in larger display spaces is beneficial.
Furthermore, immersive computational notebooks provide opportunities to better utilize VR's spatial memory~\cite{yang_virtual_2021} and multi-perspective analysis. 
For instance, users can arrange multiple code cells and visualizations side by side in 3D space, reducing the need for excessive scrolling and context-switching.
Additionally, analysts can interact with data using embodied actions~\cite{yang2020tilt,zhu2024compositingvis}, such as pointing, grabbing, or reorganizing elements dynamically, which enhance engagement and comprehension~\cite{cordeil2019iatk,huang2023embodied,yang2018origin}. 
Building on this, we identify additional opportunities to enhance performance in immersive computational notebooks through organizational strategies.
Therefore, our research aims to explore the organizational strategies within immersive computational notebook settings to better understand how analysts organize computational notebooks in immersive environments. 
We are also interested in examining how these organizational strategies vary based on the quantity of code cells.

\textbf{Organization Strategies in Immersive Environment.}
In Immersive Analytics~\cite{marriott2018immersive}, previous research has highlighted that space management~\cite{ball_move_2007, knierim2020opportunities,lin2021labeling} and content organization~\cite{liu2020design, satriadi2020maps, reiske2023multi,tong2025exploring} are pivotal in the data analysis process. 
Specifically, Liu et al. discovered that a half-cylindrical shape is the most preferred layout for multiple small data visualization blocks~\cite{liu2020design}.
In contrast, Satriadi et al. observed that a quarter-cylindrical layout was predominantly used for managing hierarchical multi-view geospatial data~\cite{satriadi2020maps}. 
Building on these findings, Liu et al. investigated how physical surroundings and co-located collaboration impact the spatial organization of virtual content in augmented reality (AR)~\cite{liu2023datadancing}. 
They identified nine distinct layouts that varied based on the physical environment, spatial structuring, and workflow dynamics.
However, these investigations mostly focused on standard digital content where sequence does not critically affect workflow. 
Few studies have explored sequenced workflows in immersive environments~\cite{in2024evaluating}, often adopting a half-cylindrical layout as their initial setup due to its efficiency in content management~\cite{liu2020design}. 
However, this perspective generally overlooks a detailed analysis of the organizational strategies that might be critical in a sequenced workflow.
Therefore, our study aims to address this gap by investigating how participants organize their views in an immersive environment using computational notebooks, where sequence significantly influences their workflow.

\textbf{Sensemaking Activities.}
Sensemaking activities are the processes where individuals and groups interpret, organize, and understand complex information to develop insights and make informed decisions~\cite{north2006toward}. 
However, previous research highlighted that sensemaking is cognitively demanding~\cite{pirolli2005sensemaking, stasko2007jigsaw, weick2005organizing}, suggesting that a well-structured organizational environment can enhance the visualizing cognitive loads~\cite{russell1993cost}. 
Andrews et al. introduced the concept of ``Space to Think'' to examine how 2D large display space supports complex sensemaking activities~\cite{andrews2010space}.
They found that spatial environments serve as external memory aids and add a semantic layer, facilitating better organization and comprehension of complex information.
Based on their successful explorations, it has been extended to immersive environments through the ``Immersive Space to Think'' (IST), which explores how 3D spaces can aid sensemaking~\cite{lisle_evaluating_2020}. 
Findings indicate that IST enhances sensemaking by offering a larger and more flexible space for document organization. 
Beyond sensemaking activities within regular text documents, sensemaking also plays a role in computational notebooks as we arrange cells in an order that makes sense to us.
Harden et al. investigated organizational strategies in 2D computational notebooks where analysts can display cells in a manner that makes sense to them~\cite{harden2022exploring}.
They found that 2D computational notebooks facilitate analysts to easily build comparative analysis and non-linear analysis workflow. 
Although sensemaking activities have been studied in regular text documents across 2D and 3D spaces, we found that exploration within computational notebooks has been limited to 2D environments.
Therefore, we are interested in exploring how immersive spaces can support the sensemaking process with the use of computational notebooks.

%% file: body/03-User_Study.tex
\section{User Study and Evaluation}
\label{sec:usability_study}
To systematically address our goals, we designed and conducted an exploratory study. 
Our primary objective was to explore organizational strategies for computational notebooks within an immersive environment where the execution order plays a significant role. 
We examined these strategies under two conditions: \simple~and \complex. 
This section details our design rationale and process, providing comprehensive information on our methodology.

\subsection{Research Questions \& Goals}
Our study aimed to address three research questions focusing on spatial structures, execution order, and the impact of increasing code cells on organizations within computational notebooks in an immersive environment.

\textbf{RQ1. What spatial structures are utilized in immersive computational notebooks?} 
We aim to identify employed spatial structures within immersive computational notebooks.
We are also interested in determining the most preferred structures.

\textbf{RQ2. What execution orders are employed in immersive computational notebooks?}
Cell placement plays a critical role in computational notebooks, as it indicates the execution order. 
Depending on how the cells are displayed, the execution order can follow either a linear or non-linear analysis workflow. 
Therefore, we aim to explore how execution orders can be visualized in an immersive space. 

\textbf{RQ3: How does an increased number of code cells impact organization strategies?}
Interacting with a larger number of code cells is inevitable as data science workflows become more complex~\cite{head2019managing}.
In such scenarios, we consider that user may focus on adopting new organizational strategies to avoid extensive navigations within massive growth in their organizations. 
Therefore, we aim to explore how the number of code cells impacts organization strategies within the immersive space.

\subsection{Experimental Setup}
\begin{figure}
    \centering
    \includegraphics[width=1\columnwidth]{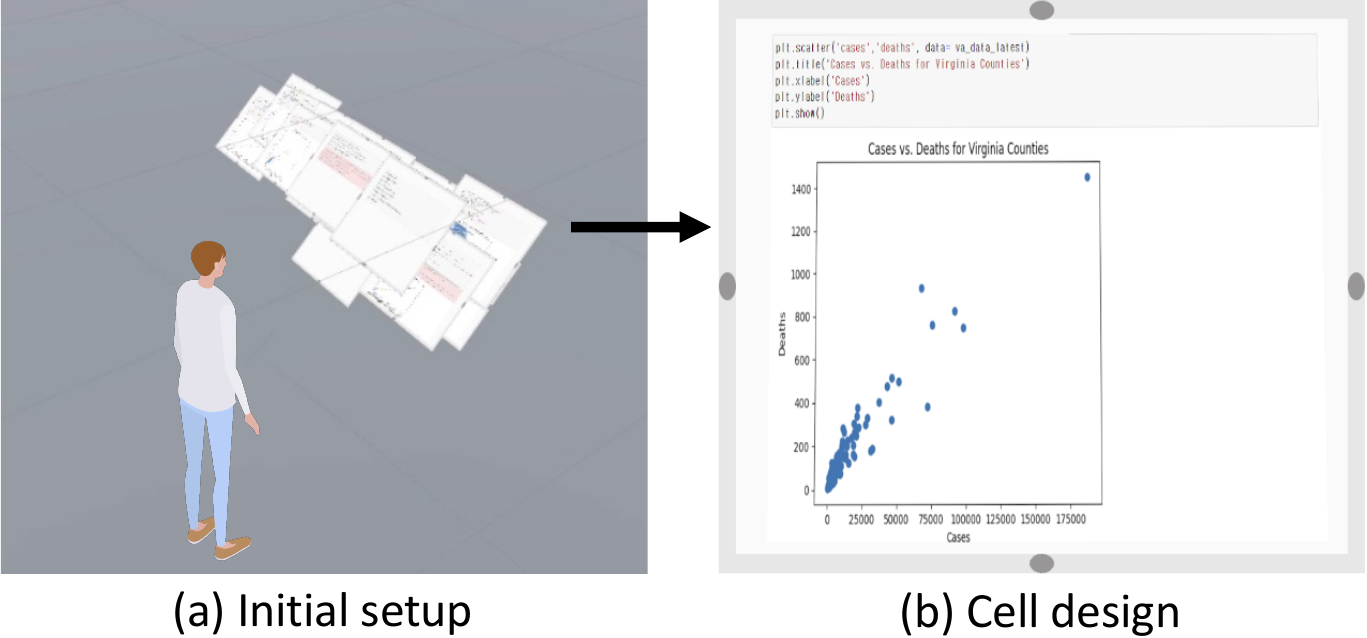}
    \caption{Illustration of the experimental setup. (a) Cells are stacked in random order. (b) A closer look at the design of the cell, where codes and outputs are located at the center. Four holes on the side are available for visualizing the execution order.}
    \label{fig:initia_setup}
\end{figure}

Building on the methodologies used in exploring 2D computational notebooks~\cite{harden2022exploring}, our study had pre-existing, non-executable code to focus on their organizational strategies rather than their programming skills.
\added{We took a screenshot of the entire Jupyter Notebook document, divided it based on each cell, and placed these cells within the immersive space.
This setup allows us to focus solely on measuring interactions related to code organization and comprehension, without introducing additional complexity such as text input for code writing or unfamiliar interfaces, like gestures or menus, for code execution.}
To enhance interaction and distinguish one cell from another, we displayed the captured cells in windows that clearly defined their boundaries, as shown in~\autoref{fig:initia_setup}-(b).
They were able to grab these windows and release them to the desired position to change the organization.
These windows included additional features that allowed participants to link cells and visually represent the execution order.
To establish these links, participants could pinch to grab one of four holes (top, bottom, left, or right) displayed on the windows and drag it to holes on other windows.

The study was conducted with the Meta Quest 2 headset, providing a resolution of $1800\times1920$.
This headset was linked to a PC with an Intel i7-11800H processor operating at 2.30 GHz and an NVIDIA GeForce RTX 3070 graphics card.
The Meta Air Link feature facilitated a cable-free experience, leveraging the PC's computation power while the headset managed the rendering. 
This configuration allowed participants to navigate a $16 m^2$ space freely.
Initially, participants were placed in the center of this space, in front of a pile of cells (see \autoref{fig:initia_setup}-(a)), each measuring $0.35x0.30m^2$, at a distance of $1m$ from them.

\subsection{Data}
We conducted the study under two conditions to deepen our understanding of the use of computational notebooks differentiated by the number of code cells, \simple~and \complex. 
To determine the appropriate number of code cells for the \simple{}, we adopted the same quantity used in a previous study on immersive computational notebooks~\cite{in2024evaluating}, \added{which included ten cells.}
For \complex, we considered various configurations. 
\added{While Luo et al. and Lisle et al. show that interacting with text-heavy content, such as during document-based sensemaking, can still be beneficial from the spatial affordances of immersive environments~\cite{lisle_evaluating_2020, luo_documents_2025}, we observed that providing an excessive number of code cells (e.g., 30-40 code cells) would make the task too challenging to finish in reasonable time.
To balance task complexity with feasibility, we selected 21 code cells by reusing code from a study by Harden et al.\cite{harden2022exploring}, which explored 2D computational notebooks and closely aligned with the objectives of our \complex.
As a result, the \complex{} involved a 133\% increase in code cells compared to the \simple{}.}
We have included the codes used in our study in the supplementary material.
The details of the code used in each condition are as follows.

\simpleB. Ten code cells were provided, each employing the K-Nearest Neighbors (KNN) classification algorithm to analyze the "iris" dataset. 
This dataset was loaded from the official scikit-learn~\cite{scikit-learn} documentation, and data visualization was conducted using matplotlib~\cite{MatplotLib} libraries. 
The visualization included two different charts for the iris dataset, each presenting the same data but created by different parameter values. 
Two parameters within the KNN methods were manipulated to determine the optimal parameter combinations.

\complexB. Twenty-one code cells were provided, focusing on the COVID-19 dataset for a state and two counties.
Participants were not required to have prior knowledge of Virginia. 
The dataset included three charts (e.g., scatter, bar, and line), each presenting the same data but analyzed differently. 
Conversely, each of the two counties was analyzed using the same methods but with its respective data, resulting in three graphs per county.
The codes used in this condition were the same as those utilized in previous studies that explored 2D computational notebooks in desktop setups~\cite{harden2022exploring}.

\begin{figure*}
    \centering
    \includegraphics[width=1\textwidth]
    {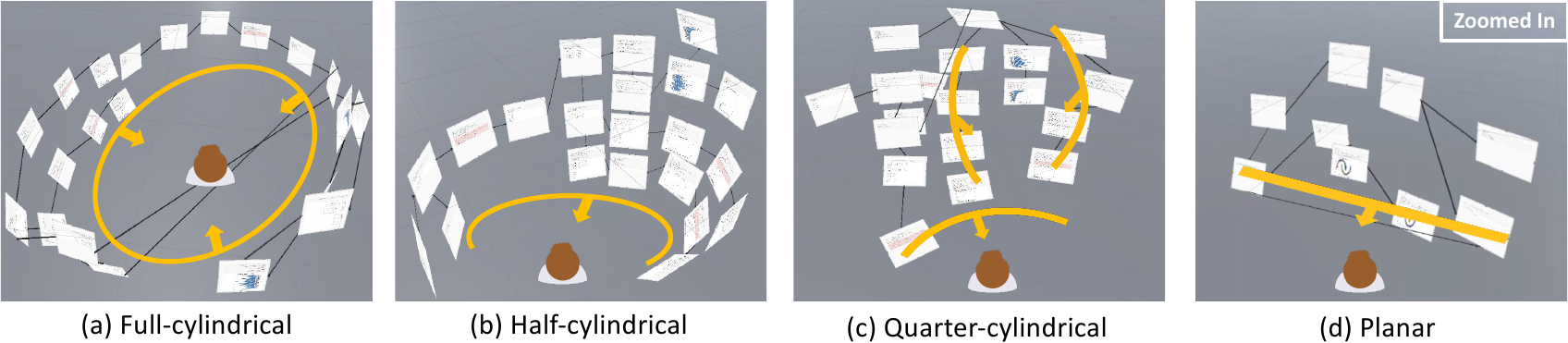}
    \caption{Implemented spatial structures. (a) Participants were fully surrounded by code cells, forming a complete cylindrical layout, whereas (b) Cells were arranged in a half-cylindrical layout, providing free space behind the participants. (c) Cells were set up only in front of the participants but still in a curved formation, forming a quarter of a cylindrical layout. (d) Cells were placed in a flat wall shape only found in the \simpleB. Note that Figure (d) has been zoomed in to show the structures clearly.}    
    \label{fig:spatial_layout}
\end{figure*}

\subsection{Participants}
We recruited 20 participants (17 male, 3 female, ages 18 to 35)
from a university mailing list.
We screened participants who had completed coursework in data science, machine learning, and computational notebooks.
Regarding VR experience, 12 participants reported having used VR, while the remaining indicated no prior experience with VR technology. 
Three of the 12 participants used VR weekly, and the remaining nine did not. 
All participants had either normal vision or vision corrected to normal.

\subsection{Procedure}
Our user study followed a full-factorial within-subjects design, with conditions balanced using a Latin square (2 groups).
The study, on average, took less than two hours.
Participants were initially welcomed and reviewed a consent form.
Then, we briefly introduced the study's objectives and procedural steps.
Following this introduction, participants proceeded to the various steps of the study as follows:

\textbf{Preparation.}
We ensured that all participants had properly adjusted their Quest 2 headsets. 
This step was essential to ensure that each participant was in comfortable conditions and could clearly see the text.

\textbf{Training.}
We began the training session by standardizing the terminology used in computational notebooks to avoid varying interpretations. 
The training was provided only once for the entire study due to the consistent operational logic across conditions, which differed only in the number of code cells. 
During the training, participants watched demonstration videos and replicated specific tasks. 
Training tasks involved methodologies similar to those used in \simple~(e.g., 10 code cells) but utilized a different algorithm, the K-Means algorithm.
They also had the opportunity to ask questions about operations or tasks. 
Recognizing the unique challenges of the VR setting, we extended the training sessions to give participants sufficient time to become familiar with the immersive environment.
This approach helped address potential VR-related issues such as physical discomfort and learning curves. 
The training concluded once participants achieved proficiency in the tasks and operations, which generally took 10-15 minutes.

\textbf{Study Task.}
After completing the training session, participants moved on to the study task. 
We re-positioned participants at the center of the room and oriented them in the same direction before starting each task to ensure consistency.
We provided a context for each task, including brief explanations of the algorithms, data, and specific tasks they were required to complete. 
Specifically, we informed that the goal is to guide new team members, enabling them to understand not only the analysis results but also the processes underlying these findings.
Additionally, we advised that the arrangement should facilitate collaboration and further development of data analysis.
To complete the given task, participants were required to reorder and link the cells in a way that made sense to them. 
Participants were not constrained by time limits for completing the tasks.

\textbf{Questionnaires.}
After each condition, participants completed a Likert-scale survey, adapted from the NASA Task Load Index (TLX), to document their subjective experiences. 
Additionally, they provided qualitative feedback on the advantages and disadvantages of the condition they had just interacted with. 
The study concluded that once participants had completed the questionnaires for all conditions.

\subsection{Measures}
In our study, we recorded the participants' entire procedures to address our research questions. 
This method allowed us to analyze not only the final layout strategies but also the intermediate steps participants took to build their final structures. 
Specifically, for RQ1, we examined how participants utilized spatial structures, and for RQ2, we analyzed the placements and orientations of their execution order in their final structures.
For RQ3, we compared \simple~and \complex~to investigate the impact of an increased number of code cells on participants' final structures.
Furthermore, this approach provides deeper insights into how participants utilize the immersive space to shape the final structures.

In addition to recording participants' processes, we collected a survey using a 7-point Likert scale to assess their subjective experiences.
Additionally, we conducted semi-structured interviews to gain deeper insights into each condition, identifying experiences and areas for improvement.

Furthermore, we collected quantitative data using specific metrics, particularly for RQ3, to investigate the impact of user interactions as the number of code cells increased.
\added{\textit{Time:} We recorded task completion time from when participants began organizing until they indicated completion of the task.}
\textit{Number of participants began with rearranging initial setups:} considering the initial setup's piled-up cells, we expected that participants might need to rearrange them into other structures to make the codes visible. 
\textit{Number of moving cells:} we recorded the total interactions involving the movement of cells for organizational purposes.
Therefore, we counted the participants who began arranging the codes before organizing them into a designated order.

\begin{table}
    \centering 
    \includegraphics[width=0.9\columnwidth]
    {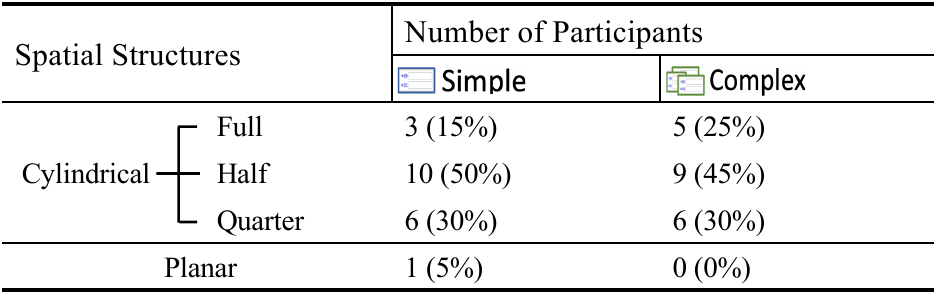}
    \caption{Distribution by spatial structures in immersive computational notebooks.}    
    \label{table:spatial}
\end{table}

%% file: body/04-Results.tex
\section{\added{Results}}
\label{sec:results}
\added{This section discusses our results.
The figures of all the final structures and the subjective ratings in \simple~and \complex~are included in the supplementary material.}

\begin{figure*}
    \centering
    \includegraphics[width=1\textwidth]{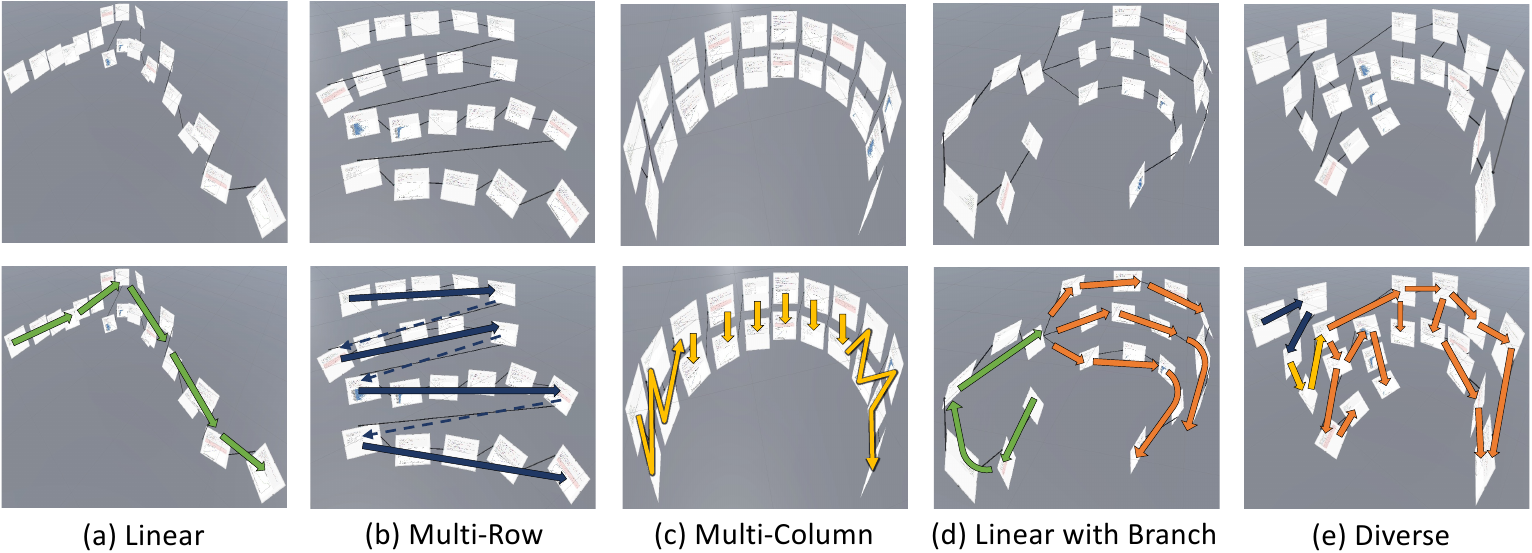}
    \caption{Illustration of execution order (top) and clarification of corresponding order (bottom). (a) Cells followed linear execution order. (b) Cells formed multiple rows, whereas (c) cells were set up in multiple columns. (d) Cells initially followed a linear order but switched to a branching strategy to facilitate code comparison. (e) Cells involved multiple execution orders simultaneously; for instance, Multi-Row, Column, and Branch were used together.}
    \label{fig:layout_everything}
\end{figure*}

\subsection{\added{Spatial Structures}}
Our study identified four primary spatial structures adopted by users: full-cylindrical, half-cylindrical, quarter-cylindrical, and planar, as illustrated in~\autoref{fig:spatial_layout}. 
We defined full-cylindrical structures when participants were fully surrounded by cells. 
Half-cylindrical structures were identified when participants arranged cells to leave free space behind them. 
Quarter-cylindrical structures were recognized when participants formed a curved cylindrical structure but left empty spaces on the back, left, and right sides. 
Lastly, planar structures were considered when cells were displayed in a flat wall shape without any curvature.
We observed that the majority of participants chose half-cylindrical structures, with 50\% in \simple~and 45\% in \complex. 
Quarter-cylindrical structures were the second most popular, selected by 30\% of participants in both conditions. 
While half and quarter-cylindrical structures were consistently used across conditions, other structures varied: 15\% of participants in \simple~adopted full-cylindrical structures, and the planar structure was the least utilized, with only 5\%.
However, with \complex, observation shifted: 25\% of participants opted for full-cylindrical structures, and notably, no participants chose planar structures (see~\autoref{table:spatial}).

\subsection{\added{Execution Orders}}
Across both conditions, our study identified five primary execution orders that were adopted by users: linear, multi-row, multi-column, linear with branch, and diverse, as illustrated in~\autoref{fig:layout_everything}. 
We defined a linear execution order as being laid out in a single direction, either left-to-right, top-to-bottom, or in the reverse direction.
However, a non-linear execution order was defined when multiple directions were employed, or directions were split into different paths. 
A multi-row execution order was defined when changes in execution order created additional rows, while a multi-column execution order was identified when changes created additional columns.
In addition, we defined linear with a branch where multiple changes in execution order were made from a single cell. 
Lastly, we categorized an execution order as diverse when participants employed more than two types of execution orders.
While only 20\% in \simple~and 15\% in \complex~utilized linear execution order, a majority of participants--80\% in \simple~and 95\% in \complex--chose non-linear execution order.
This non-linearity included linear with branch, multi-row, multi-column, and diverse execution order (refer to~\autoref{table:execution_order} for details).

\begin{table}
    \centering
    \includegraphics[width=0.85\columnwidth]{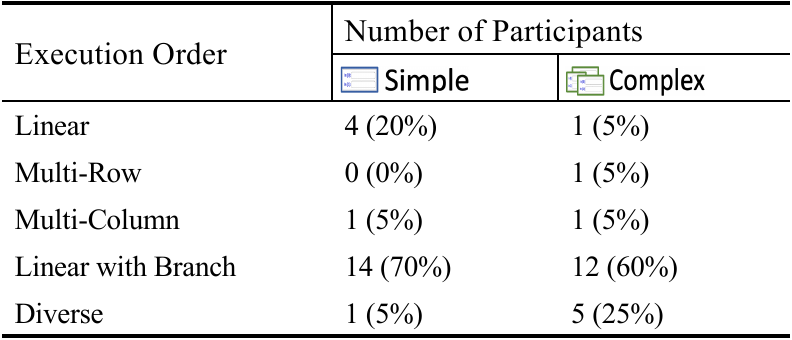}
    \caption{Distribution of execution order in immersive computational notebooks.}
    \label{table:execution_order}
\end{table}

\subsection{\added{Quantatitive Results}}
This section reports statistical analyses of both performance metrics and subjective ratings. 
For continuous dependent variables that met normality assumptions, we applied linear mixed-effects models with random intercepts~\cite{Bates2015}. 
Pairwise comparisons were conducted using Tukey-adjusted contrasts on estimated marginal means~\cite{Lenth2016}, with standard errors (SE), $z$-values, and $p$-values reported.

For ordinal data or non-normally distributed measures, we applied the Wilcoxon signed-rank test for within-subject pairwise comparisons~\cite{wilcoxon1945individual}. 
We report the test statistic ($V$), standardized $Z$ value, $p$-value, and the effect size $r$, computed as $r = \frac{Z}{\sqrt{N}}$, where $N$ is the number of participants~\cite{fritz2012effect}. 
Furthermore, we use the following notation for statistical significance: *** for $p < 0.001$, ** for $p < 0.01$, and * for $p < 0.05$.
Mean values are presented with 95\% confidence intervals (CIs). 
Complete statistical results are available in the supplementary materials.

\textbf{Time}: 
We first measured the overall task completion time for each condition, followed by a breakdown based on the observed structures and execution orders. 
A linear mixed-effects model with a random intercept for participant revealed a significant main effect of condition ($\beta = –1.33$, $SE = 0.18$, $\chi^2(1) = 31.88$, $p < 0.0001$).
Pairwise comparison also confirmed that completion time was significantly greater ($\beta = -1.33$, $SE = 0.18$, $z(40) = –7.22$, $p < 0.0001$) for the \complex~($avg. 1263.17s$, $CI= 238.18s$) than for the \simple~($avg. 514.94s$, $CI= 109.23s$), as shown in Fig.\ref{fig:time}-(a).

\textbf{Number of participants began with rearranging initial setups}:
In the initial setup, all code cells were stacked, making them not fully visible. 
As a result, participants needed to rearrange the cells to view the entire code set. 
We recorded whether participants began by reorganizing the initial layout. 
In the \simple, 85\% (17 out of 20, $CI= 0.17$) of participants initiated the task by reorganizing the layout, compared to 65\% (13 out of 20, $CI= 0.23$) in the \complex~(Fig.\ref{fig:time}-(b)).
A linear mixed-effects model with a random intercept for participant revealed no significant effect of condition ($\beta = 0.2$, $SE = 0.12$, $\chi^2(1) = 2.86$, $p < 0.091$).
Pairwise comparison also showed a non-significant difference between conditions ($\beta = 0.2$, $SE = 0.12$, $z(40) = 1.71$, $p < 0.087$).

\textbf{Number of Moving cells}:
We counted each time participants changed the position of individual cells.
Furthermore, we break down the results based on whether participants began by rearranging cells or not.
Our analysis showed that the number of cell movements was significantly affected by the number of code cells (***), with a marginal effect observed based on the participant beginning by rearranging cells (*).
A linear mixed-effects model with a random intercept for participant revealed a significant main effect of condition ($\beta = –1.80$, $SE = 0.16$, $\chi^2(1) = 52.34$, $p < 0.0001$).
Pairwise comparisons indicated that the number of movements was significantly greater ($\beta = -1.80$, $SE = 0.16$, $z(40) = –11.36$, $p < 0.0001$) for the \complex~($avg. 171.2$, $CI= 35.45$) than for the \simple~($avg. 47.6$, $CI= 7.68$), as shown in Fig.\ref{fig:time}-(a).

Furthermore, participants who started by rearranging cells moved an average of 49 times, and an average of 31 times for those who did not rearrange in \simple~(see Fig.\ref{fig:time}-(d)-simple). 
Notably, this observation shifted in \complex: those who started by rearranging cells had an average of 148 interactions; however, this number dramatically increased with those who did not rearrange cells by 45\% to an average of 215 interactions (see Fig.\ref{fig:time}-(e)-complex).

\textbf{Ratings and Ranking.}
We observed marginal effects in mental demand (*), physical demand (*), and perceived effort (*). 
Specifically, average ratings for mental demand were 2.80 for \simple~and 3.55 for \complex, physical demand were 2.25 for \simple~and 3.20 for \complex, and perceived effort were 1.95 for \simple~and 2.90 for \complex. 
However, frustration did not differ significantly between conditions (3.85 for \simple, 4.15 for \complex).

%% file: body/05-Discussion.tex
\begin{figure}
    \centering
    \includegraphics[width=1\columnwidth]{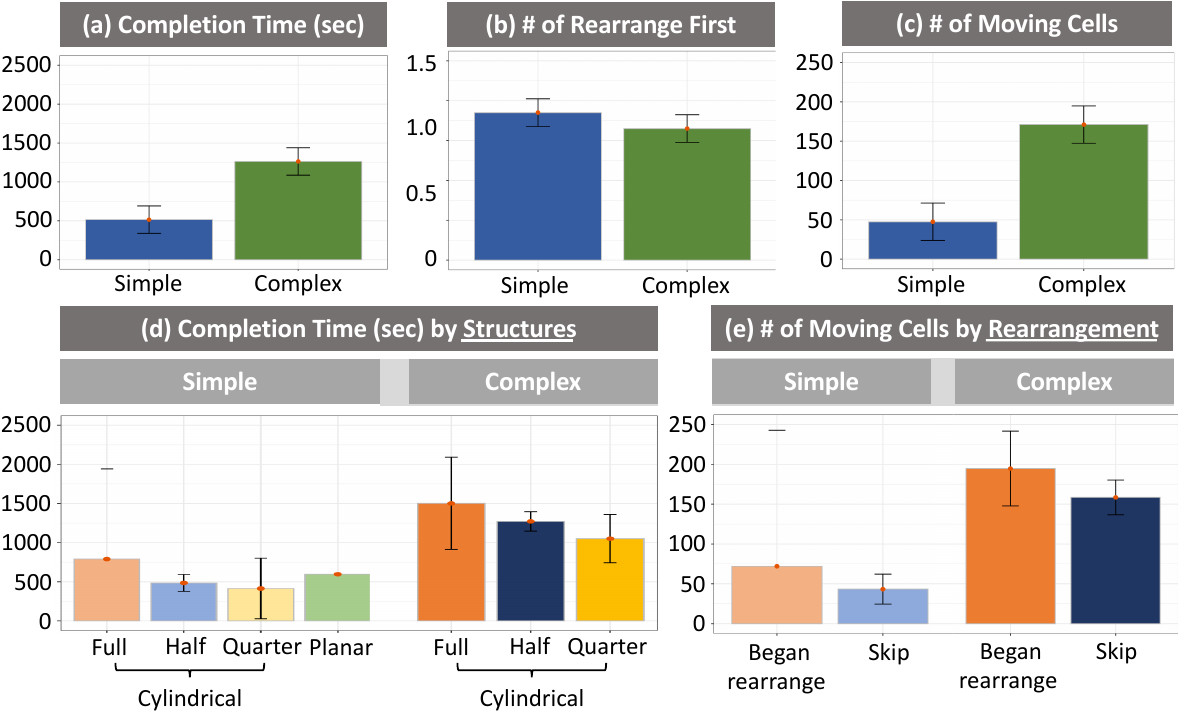}
    \caption{\added{Collected quantitative data on (a) task completion time, (b) the number of participants who began by rearranging the initial setup, and (c) the number of cell movements. Additionally, (d) includes a breakdown of completion time by the spatial structures constructed, and (e) categorizes the number of cell movements based on whether participants started with rearrangement.}}
    \label{fig:time}
\end{figure}

\section{Key Findings \& Discussion}
\label{sec:discussion}

\begin{figure}
    \centering
    \includegraphics[width=1\columnwidth]{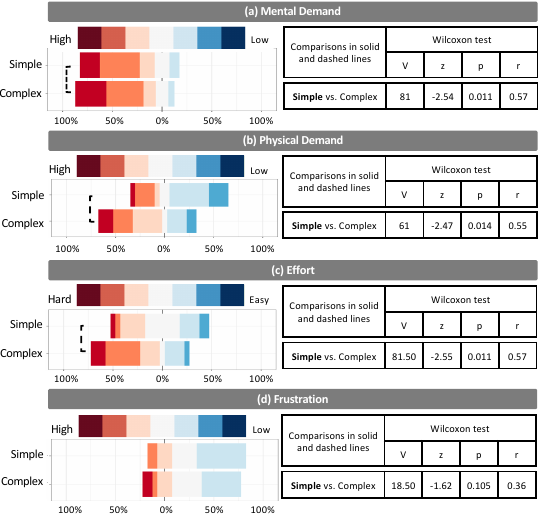}
    \caption{\added{Participants' subjective ratings: (a) mental demand, (b) physical demand, (c) effort, (d) frustration level. The tables display the Wilcoxon signed-rank test involves a test statistic ($V$), the standardized $z$ value, $p$-value, and the effect size $r$. Additionally, dashed lines indicate statistical significance with p$<$0.05.}}
    \label{fig:ratings}
\end{figure}

\added{\textbf{Half-cylindrical structure largely favored by participants (RQ1).}}
Our findings show that participants predominantly chose half-cylindrical structures over other structures when organizing computational notebooks in immersive environments. 
\added{We believe this preference is potentially due to the clearer separation of the starting and ending points of an analysis through visible distance, which participants found makes more sense than the continuous flow in full-cylindrical structures.
At the same time, the curvature of half-cylindrical structures provides a stronger sense of immersion than quarter-cylindrical or planar structures \cite{liu2022effects}.
We believe this observation aligns with findings by Satriadi et al.~\cite{satriadi2020maps}, who suggested a structural preference when engaging with small multiples and employing physical interactions in immersive environments.
In summary, our findings extend existing spatial organization taxonomies~\cite{andrews2010space, liu2020design, luo_documents_2025}—shifting the focus from general documents to code-centric artifacts.}

\added{\textbf{Task completion times tended to increase as the spatial structures became more enclosed (RQ1).}
While no substantial differences in execution order were observed, spatial structure appeared to influence time performance. 
Specifically, participants who constructed more enclosing structures tended to take longer compared to those who built less enclosing structures (see Fig.\ref{fig:time}-(a)).
Although previous studies suggest that participants tend to avoid effort-intensive tasks \cite{batch2019there}, our findings reveal a consistent preference for half-cylindrical structures over quarter-cylindrical ones, even though the latter were faster to build. 
Such behavior aligns with findings by Andrews et al.~\cite{andrews2010space}, who showed that users are often willing to invest additional effort into organizing artifacts when it enhances interpretability and supports the externalization of evolving mental models.}

\added{\textbf{Immersive computational notebooks potentially support greater adoption of non-linear analysis compared to desktop environments (RQ2).}
Although our study did not include a direct comparison to desktop-based notebooks, we used the same dataset as prior work in 2D environments~\cite{harden2022exploring}, enabling an indirect comparison.
They found that freeform organization encourages participants to externalize their perceived execution order in ways that better align with their reasoning processes, often by utilizing non-linear execution orders.
Interestingly, in our study, the use of non-linear execution orders increased by approximately 30\% compared to those observed in 2D notebook settings. 
We believe this shift is driven by the embodied interaction enabled by VR, which allows participants to more naturally externalize their reasoning processes~\cite{in2024evaluating, in2023table}.
Consequently, we posit that immersive computational notebooks potentially allow users to build and organize execution orders in ways that more intuitively reflect their analytical reasoning.}

\begin{figure}
    \centering
    \includegraphics[width=1\columnwidth]{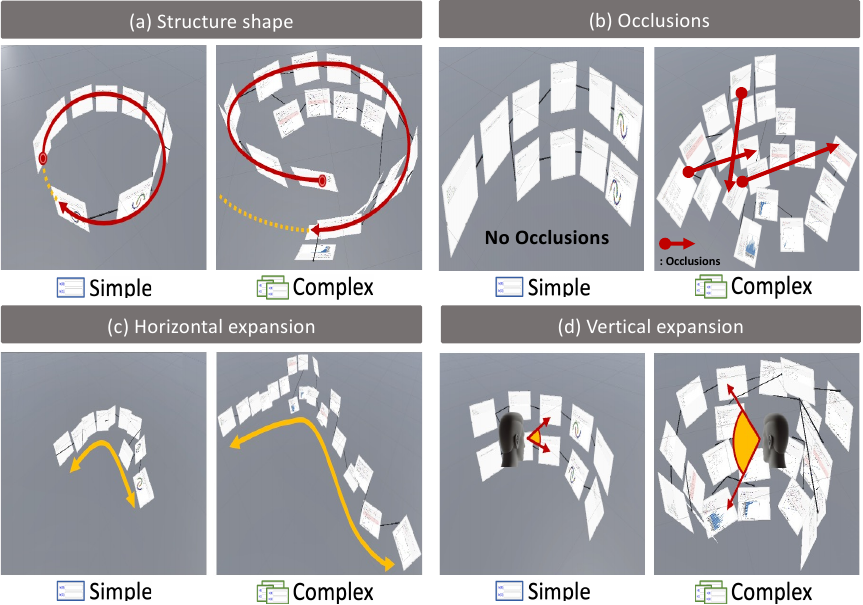}
    \caption{Illustration of the impact of increasing the number of code cells on organizations. (a) The full-cylindrical structure has a spiral shape in \complex, while the structure is circular in \simple. (b) The employed links in \complex occlude notebooks, while such occlusions were not evident in \simple. (c) illustrates horizontal expansion, and (d) illustrates vertical expansion as the number of cells increases.}
    \label{fig:number_cell}
\end{figure}

\textbf{The functions of links in execution order (RQ2).}
While analyzing the employed execution orders, we observed that the linking mechanism used in our study could be classified into two distinct types: short and long links.
Shorter links primarily functioned to indicate the typical sequencing orders, often following a linear execution order such as top-to-bottom, left-to-right, and reverse counterparts. 
In contrast, longer links were employed to either initiate a new execution order or continue an existing order in a different space. 
For instance, in \autoref{fig:layout_everything}-(b), a participant used longer links to establish multiple rows of execution, while shorter links maintained sequential connections within those rows.
Similarly, \complex~in \autoref{fig:number_cell}-(b), longer links (on the far right) were used to start a new branch in a separate space.
Interestingly, we noticed that all participants who employed diverse execution orders used longer links to visualize their execution order in \simple~and \complex. 
Therefore, we consider that diverse execution orders in our tested task potentially result in more spatially distributed structures, requiring longer links to maintain logical connections between distant cells.

\begin{figure}
    \centering
    \includegraphics[width=0.85\columnwidth]{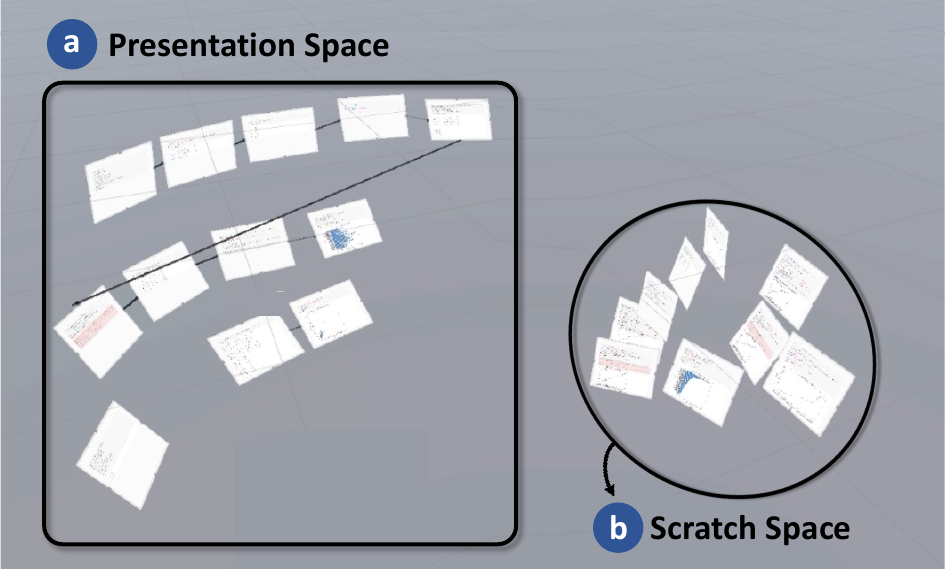}
    \caption{Illustration of space management strategies within immersive computational notebooks. (a) Participants organized the cells in a structured manner, whereas in (b), the placement of cells appeared less organized.}
    \label{fig:space_usage}
\end{figure}

\textbf{Space management strategies in building final organizations (RQ1 and RQ2).}
\added{We identified that the sensemaking activities identified in prior studies extend to the use of immersive computational notebooks, where users create two different spatial zones to support the development of emerging insights \cite{ball2008effects, cavallo2019dataspace, lisle_evaluating_2020}.
In the context of computational notebooks, we observed that the majority of participants (75\%, 15 out of 20) divided the given space into presentation and scratch spaces in the process of building final structures (see ~\autoref{fig:space_usage}).}
In the presentation space, participants displayed their final structures with deliberately arranged execution orders, indicating that this area is for refining presentations and reports derived from their analyses.
By isolating presentation space from others, participants were able to focus more effectively~\cite{chen2021effect} on organizing their final structures.
In the scratch space, which served as a temporary holding area, participants placed code snippets that were not in sequenced order. 
Utilizing this space enabled participants to keep code snippets accessible for review and integration at a later stage.
More specifically, within the scratch space, participants further divided this area for various purposes. 
60\% of participants (9 out of 15) grouped similar codes with corresponding visualizations, while 40\% (6 out of 15) organized code cells and visualizations separately. 
As one participant explained, ``Clustering code cells enables comparative analysis easier (P7)''.

\textbf{Increased number of cells impacted the organizational strategies (RQ3).} 
As the number of code cells increased, we observed notable structural changes. 
Two participants shifted to a full-cylindrical structure—one from a half-cylinder, another from a planar structure.
This shift likely occurred because the full-cylindrical structure can accommodate more artifacts within a curved shape~\cite{chen2021urbanrama}.
Interestingly, unlike in the \simple, maintaining a circular shape became difficult in \complex, resulting in a spiral shape, as shown in~\autoref{fig:number_cell}-(a). 
This spiral shape, however, required participants to walk through other cells to access those positioned behind, making interactions less intuitive.
Additionally, structures grew significantly—horizontally or vertically (see \autoref{fig:number_cell}-(c) and (d))-introducing physical demands in navigation \cite{kuber2023alterations}.
\added{This aligns with our findings: participants reported higher physical demand in \complex~(2.7 out of 5) compared to \simple~(2 out of 5), a 25\% increase (see Fig.\ref{fig:ratings}-(b)). 
One participant noted, ``Navigation was dramatically harder (P11)''. 
We believe that the considerable time spent on extensive navigation partially contributed to participants' longer task completion times (see Fig.\ref{fig:time}-(a)).}

Regarding execution orders, we observed that 35\% (7 out of 20) changed their execution order as the number of code cells increased.
Among those, we found more than half of the participants (4 out of 7) changed their execution order to diverse in \complex, where they all had linear with a branch in \simple.
Furthermore, we identified that 10 participants utilized longer links to visualize execution order in the \complex, whereas only two participants used longer links in the \simple. 
We believe this is due to the difficulty of finding suitable space around the user within expanded structures to start a new execution order. 
For instance, participants may have found that the only available space was on the ceiling or floor, leading to further vertical expansion, which potentially made navigation more cumbersome~\cite{tseng2023understanding}.
As a result, we believe participants placed those cells in a completely different space instead.
Moreover, among participants who utilized longer links, nearly all (9 out of 10) had links that passed through or over other cells, occluding critical information (see \autoref{fig:number_cell}-(b)).

\textbf{The impacts on user interactions when the number of code cells increased (RQ3).}
In addition to the impact of the increased number of code cells on organizational strategies, it also had a notable impact on user interactions.
In the process of shaping their organizations, participants frequently needed to revisit cells, making it inevitable for them to memorize the spatial placement of previously visited cells to locate them efficiently.
Interestingly, identifying revisited cells proved more challenging in \complex; however, this issue did not arise in \simple.
We believe that the benefits of spatial information in boosting memory capabilities~\cite{lu2025ego, yang_virtual_2021, zagermann2017memory} are diminished when participants interact with a larger number of cells.
\added{This was partially supported by our findings: the average mental demand rating in the \complex~condition increased by 22\% (see Fig.\ref{fig:ratings}-(a)), and 11 participants reported related challenges. 
For example, one participant noted, ``Identifying specific cells was challenging, as I needed to recall many cells (P6)''.}

Challenges further arose as the number of code cells increased.
We observed that a growing number of participants began organizing the code into the final structure without a comprehensive understanding of the given codes in \complex.
To gain deeper insights into this observation, we counted the number of interactions associated with cell movements. \added{Participants who began by rearranging cells performed fewer interactions in both the \simple~and \complex. 
We believe that participants considered rearranging cells to be worth nothing to their reasoning process, and thus were not motivated to invest additional effort in the organizational process.}
Although participants believed that skipping the step of rearranging cells would reduce effort, it ultimately resulted in increased efforts and interactions.
This observation is confirmed by our ratings of ``effort'' in the NASA TLX survey, with an average score of 3.16 out of 5 for those who did not arrange first, slightly higher than the score of 2.69 out of 5 for those who arranged initially.

%% file: body/06-Implication.tex
\section{\added{Design Implications}}
\label{sec:implication}

\added{\textbf{DI-1. Provide half-cylindrical structures as the initial setup.}
Analysts should carefully consider their initial organization in immersive environments, since it can serve as a spatial anchor that significantly influences organizational strategies in later stages of analysis~\cite{luo2022should}.
Participants in our study predominantly adopted half-cylindrical structures, which offered a balance between immersion and visual clarity. 
In addition, Enriquez et al.~\cite{enriquez2024evaluating} demonstrated the effectiveness of a half-cylindrical structure in supporting transitions from desktop to immersive environments. 
This suggests that a half-cylindrical structure may also be beneficial in broader use cases where analysts transition from desktop-based to immersive notebooks.
As a result, we suggest adopting a half-cylindrical structure as the initial setup for immersive computational notebooks.}

\added{\textbf{DI-2. Support for non-linear execution order while avoiding visual clutter.}
Our study suggests that non-linear execution orders—such as branches, multi-row/column, and diverse—more closely reflect analysts’ perceived workflows~\cite{weinman2021fork}. 
However, in immersive environments, participants often used longer links to represent these complex execution paths.
These links often caused occlusion or crossed through their workspace, potentially leading to confusion or a loss of context. 
To address this, analysts should be able to control the visibility of execution links—for example, by toggling their display or showing only the links relative to a selected cell—to reduce visual clutter and maintain clarity during analysis.}

\added{\textbf{DI-3. Utilize semantic zoning.}
The large space available in immersive environments can be divided into multiple zones and assigned meaningful functions~\cite{lisle_evaluating_2020}. 
In our study, we observed participants organizing their workspace into multiple zones—for example, using one as a scratch space for experimentation and another as a presentation space for refined outputs. 
This concept can be extended to enhance data exploration capabilities, particularly given that current immersive notebook systems often lack intuitive data exploration, such as data transformation and visualization.
For instance, one zone could serve as a code-focused workspace for hypothesis testing and detailed analysis, while another could function as an embodied data exploration area, enabling users to interact with data through gestures rather than code.
This implication may even broaden accessibility for analysts with entry-level coding experience~\cite{in2023table}.
}

\added{\textbf{DI-4. Support low-interaction organizational processes for large-scale analysis.}
As the number of code cells increases, the size of the workspace expands significantly, requiring extensive navigation.
In our study, analysts had to consider the trade-off between interaction efficiency and organizational clarity.
However, an increasing number of participants hesitated to reorganize code cells due to the anticipated extensive efforts, as they needed to interact with individual cells one at a time.
One potential solution is to enable users to manipulate groups of cells, thereby reducing repetitive interactions. 
A further approach is to provide the templates, similar to pre-defined layouts in tools like Miro~\cite{miro} and Figma~\cite{figma}.
This implication would help users to organize quickly and with minimal interaction.}

%% file: body/07-Limitation.tex
\section{Limitation \& Future Work}
\label{sec:limitation}
While we identified various spatial structures, execution orders, and their preferences within immersive computational notebooks, our study leaves several areas open for further exploration.
In our tested task, half-cylindrical structures remained dominant, indicating that the workspace constraints observed in previous studies persist~\cite{in2024evaluating}.
We believe this is due to the reliance on physical-based interaction methodologies, which inherently influence how participants organize their workspace (i.e., users may place cells around them to keep content within an easily reachable distance).
We consider that interaction techniques commonly used for analyzing visualizations could be adapted in immersive computational notebooks, as they enable users to identify information with fewer interactions and from a far distance.
We envision it could allow participants to build larger half-cylindrical structures while still maintaining accessibility and efficient navigation.
Specifically, methods such as focus+context using various types of lenses (e.g., Cartesian and fisheye lenses)~\cite{yang2022pattern}, overview+detail, and pan-and-zoom techniques~\cite{yang2020embodied} may result in shaping different spatial structures in the use of immersive computational notebooks.

Furthermore, we provided divided code cells from a single document; however, we realized during our literature that real-world data analysis often involves utilizing multiple documents simultaneously.
To better mirror real-world analysis, we considered that incorporating code cells from multiple documents needs to be explored to further deepen understanding of the organizational strategies. 
Additionally, computational notebooks are commonly used for collaboration~\cite{chattopadhyay2020s}.
We noted that collaboration substantially influences organizational strategies~\cite{enriquez2024evaluating,mendez2019toward,tong2023towards}, as the process of understanding and structuring complex information can be adjusted by others. 
Recognizing this, we believe that future work can explore the organization's strategies when multiple data analysts collaborate within immersive computational notebooks.

%% file: body/08-Conclusion.tex
\section{Conclusion}
\label{sec:conclusion}
In this study, we aim to deepen our understanding of the spatial structures employed for computational notebooks and examine how various execution orders can be visualized in an immersive context. 
Our findings indicate that the half-cylindrical structure is predominantly used, and a linear with branching execution order was mostly chosen by participants. 
Notably, we observed a significant increase in the adoption of non-linear analysis in immersive computational notebooks. 
This observation underscores the potential of computational notebooks in immersive spaces to encourage analysts to engage more in non-linear analysis. 
However, as the number of cells increased, structures expanded, and the tendency to adopt multiple execution orders increased.
We identified that navigating these expanded structures and managing multiple execution orders with the current physical-based interaction methodologies used in immersive computational notebooks may introduce navigational challenges. 
As a result, alternative interaction methodologies should be explored to enhance the organization of cells and maintain efficiency when working with larger cells.